\documentclass{article}
\usepackage{authblk}
\usepackage{amsmath, amssymb}
\usepackage{graphicx}
\usepackage{hyperref}
\usepackage{graphics}
\usepackage{rotating}
\usepackage{tikz-feynman}
\usetikzlibrary{feynman}

\begin{document}

\title{Electromagnetic and weak decay of singly Heavy Baryons (Qqq)}

\author{Kinjal Patel \thanks{kinjal1999patel@gmail.com}}

\author{Kaushal Thakkar\thanks{Corresponding Author: kaushal2physics@gmail.com}}

\affil{Department of Physics, Government College, Daman-396210, \\
U. T. of Dadra $\&$ Nagar Haveli and Daman $\&$ Diu, Veer Narmad South Gujarat University, Surat, India }

\maketitle
\begin{abstract}
The heavy-to-heavy exclusive semileptonic transitions of singly heavy baryons (SHBs) are investigated within the framework of the Hypercentral Constituent Quark Model (hCQM). The six-dimensional hyperradial Schr\"{o}dinger equation is solved in the variational approach to calculate the ground state masses of bottom and charmed baryons. The transition magnetic moments and radiative $M1$ decay widths are calculated using the spin-flavour wave function and the effective quark masses of constituent baryon. The Isgur-Wise function (IWF) is determined at zero recoil to compute the $b \rightarrow c$ semileptonic decay. Additionally, the branching ratios, as well as the slope and convexity parameters of IWF are evaluated and compared with results from other studies.
\end{abstract}

\section{Introduction}\label{sec1}

Significant experimental advancements have been made in the past few years in the investigation of heavy baryon characteristics. Experiments have successfully observed nearly all $1S$ state of baryons containing a single heavy quark. Additionally, several $1P$ state of singly heavy baryons have been detected by BABAR, Belle and LHCb. As of now, approximately 32 charmed and 30 bottom-flavoured singly heavy baryons are listed in the Particle Data Group (PDG) \cite{pdg2024}. The $\Lambda_c^+$ was first reported by Fermilab in 1976 \cite{Knapp1976}. The $\Lambda_b^0$ baryon was discovered early at CERN ISR and later reported by several collaborations \cite{{Bari1991a},{Bari1991b}}. The two states $\Sigma_c^{+*}$ and $\Sigma_c^+$ were observed in their $\Lambda^+_{c}\pi^0$ decay \cite{Ammar2001}. The excited bottom-strange states $\Xi_b(6327)^0$, $\Xi_b(6333)^0$, $\Xi_b(6100)^-$ and $\Xi_b(6227)^0$ have been reported by LHCb and CMS collaboration \cite{Aaij1,Aaij2,Sirunya2021}. $\Sigma_b^0$, $\Sigma_b^{0*}$ and $\Omega_b^{-*}$ baryons are not confirmed yet. Over the years, the study of heavy baryons has been motivated by advances in both experimental facilities and theoretical models. A singly heavy baryon ($Qqq$) consists of one heavy quark $Q = b, c$ and two light quarks $q = u, d$ and $s$. The heavy $c, b$ quarks are much heavier than the Quantum Chromo Dynamics (QCD) scale, whereas the light $u, d, s$ quarks are much lighter than $\Lambda_{QCD}$: $m_c, m_b \gg \Lambda_{QCD} \gg m_u, m_d, m_s$. This separation of mass scales allows the heavy quark to act as a static colour source, decoupling its dynamics from those of the light quarks. In the heavy mass limit, the heavy quark's spin and orientation have minimal influence on the overall baryonic properties.
\\

In the study of heavy baryons, magnetic moments, radiative decays and semileptonic decay are very important. Crucial information on the spin structure and dynamics of quarks in the baryon is provided by the magnetic moments. Magnetic moments provide a window into the spatial distribution and alignment of quark spins within baryons. For heavy baryons, the dominant contribution to the magnetic moment arises from the light quarks because the heavy quark provides comparatively small magnetic moment. This distinction highlights the interplay between the heavy and light quarks. Radiative decay, characterised by the emission of a photon during transitions, serve as a powerful probe of the electromagnetic structure and interactions of heavy baryons. The radiative decay width is proportional to the cube of the photon momentum and the masses of the initial and final states of the baryon. Thus, the radiative decays provides a valuable tool for studying the energy scales and mass hierarchies in baryons. The precise measurements of semileptonic decays can help determine the Cabibbo-Kobayashi-Maskawa (CKM) matrix elements, which are fundamental to understanding quark flavour transitions in the Standard Model. The decay mediated by the weak interaction involves the transition of a heavy quark ($b \rightarrow c$) accompanied by the emission of a lepton and an anti-neutrino. The theoretical description of semileptonic decay relies heavily on the Isgur-Wise function (IWF).\\

Singly heavy baryons have been extensively studied using a variety of theoretical approaches, including the Quark-Diquark Model \cite{Nejad2020}, Bethe-Salpeter Equation \cite{Ivanov1999}, QCD Sum Rules \cite{Wang2020}, Lattice QCD \cite{Brown2014}, Relativistic Quark Model \cite{{Faustov2016},{Ebert2006}}, Chiral Perturbation Theory \cite{Wang2019}, Light Front Approach \cite{Hong-Wei2008}, Light-Cone Sum Rules \cite{Luo2024}, and the Bag Model \cite{Bernotas2013}. Various properties of SHBs have been investigated, including mass spectra \cite{Kakadiya2022,{Kim2018},{Wei2017},{Weng2024}}, magnetic moments and radiative decays \cite{Hazra2021}, \cite{Wang2019}, \cite{Chen2024}, semileptonic decays \cite{{Farhadi2023},{Thakkar2024}} and lifetimes \cite{Cheng2023}. Among these, the $\Lambda_b^0$ and $\Lambda_c^+$ baryons are the most thoroughly studied, with extensive analysis reported in \cite{{Colangelo2020},{Thakkar2020},{Guo1996},{Galkin2020},{Dai1996},{Gutsche2015},{Efimov1992}}. The $\Lambda_b^0$ and $\Lambda_c^+$ baryons are also experimentally well established which allows testing of various theoretical models. Experimental status and quark content for bottom and charmed baryons are shown in Table \ref{tab:table1}. The status is listed as poor ($*$), only fair ($**$), very likely to be certain ($***$) and certain ($****$) for their existence.\\

In this paper, we have extended our previous work \cite{Thakkar2020} on the exclusive semileptonic decay of $\Lambda_b^0$ baryon. Here, we study the ground state masses, magnetic moments, transition magnetic moments, radiative decay and semileptonic decay of singly heavy baryons. This paper is organised as follows: In section \ref{sec:1}, we calculate the ground state masses of the singly heavy baryons in hCQM. The magnetic moments and radiative decays are discussed in Section \ref{sec:2}. The Isgur-Wise function (IWF) and $b \rightarrow c$ semileptonic decay widths are computed in section \ref{sec:3}. The result is presented and discussed in Section \ref{sec:4}. The paper is summarised in Section \ref{sec:5}.

\begin{table*}
\begin{center}
    \caption{{\label{tab:table1} Quark content and experimental status of singly Heavy Baryons}}
    \centering
    \begin{tabular}{ccccc}
    \hline
    Baryon	&	Quark	&	Experimental & Baryon		&	Experimental 	\\
	$J^P=\frac{1}{2}^+$&	Content	&	Status \cite{pdg2024}	 &$J^P=\frac{3}{2}^+$	 &Status \cite{pdg2024}	\\
    \hline								
    $\Sigma_b^+$	&	uub	&	***	&	$\Sigma_b^{+*}$	&	***	\\
    $\Sigma_b^0$	&	udb	&	-	&	$\Sigma_b^{0*}$	&	-	\\
    $\Sigma_b^-$	&	ddb	&	***	&	$\Sigma_b^{-*}$	&	***	\\
    $\Lambda_b^0$	&	udb	&	***	&	$\Lambda_b^{0*}$	&	***	\\
    $\Xi_b^0$	&	usb	&	***	&	$\Xi_b^{0*}$	&	***	\\
    $\Xi_b^-$	&	dsb	&	***	&	$\Xi_b^{-*}$	&	***	\\
    $\Omega_b^-$	&	ssb	&	***	&	$\Omega_b^-$	&	-	\\
    $\Sigma_c^{++}$	&	uuc	&	****	&	$\Sigma_c^{++*}$	&	***	\\
    $\Sigma_c^{+}$	&	udc	&	****	&	$\Sigma_c^{+*}$	&	***	\\
    $\Sigma_c^{0}$	&	ddc	&	****	&	$\Sigma_c^{0*}$	&	***	\\
    $\Lambda_c^{+}$	&	udc	&	****	&	$\Lambda_c^{+^*}$	&	***	\\
    $\Xi_c^{+}$	&	usc	&	***	&	$\Xi_c^{+*}$	&	***	\\
    $\Xi_c^{0}$	&	dsc	&	****	&	$\Xi_c^{0*}$	&	***	\\
    $\Omega_c^{0}$	&	ssc	&	***	&	$\Omega_c^{0*}$	&	***	\\
    \hline
    \end{tabular}
\end{center}
\end{table*}

\section{Theoretical Framework}\label{sec:1}
The properties of singly heavy baryons are studied within the framework of Hypercentral constituent quark model (hCQM). Which is well established model and effective in describing the internal dynamics and properties of baryons \cite{{Ferraris1995},{Giannini1983}}. In hCQM, Jacobi coordinates are essential for simplifying the three-body problem, providing a simplified representation of inter-quark dynamics. Jacobi coordinates provides the relevant degrees of freedom for the relative motion of the three constituent quarks and are given as
\begin{equation}\label{eq:1}
\boldsymbol{\rho} = \frac{1}{\sqrt{2}}(\mathbf{r}_1 - \mathbf{r}_2)
\end{equation}
\begin{equation}\label{eq:2}
\boldsymbol{\lambda} = \frac{{m_1 \mathbf{r}_1 + m_2 \mathbf{r}_2 - (m_1 + m_2) \mathbf{r}_3}}{{\sqrt{m_1^2 + m_2^2 + (m_1 + m_2)^2}}}
\end{equation}
The reduced masses are given as
\begin{equation}\label{eq:3}
m_{\rho}=\frac{2 m_{1} m_{2}}{m_{1}+ m_{2}}
\end{equation}
\begin{equation}\label{eq:4}
m_{\lambda}=\frac{2 m_{3} (m_{1}^2 + m_{2}^2+m_1m_2)}{(m_1+m_2)(m_{1}+ m_{2}+ m_{3})}
\end{equation}
Here $m_1$, $m_2$ and $m_3$ are the constituent quark masses.
The kinetic energy operator in the center of mass frame ($R_{c.m.} = 0$) can be written as
\begin{equation}\label{eq:5}
\frac{P_x^2}{2m} = -\frac{\hbar^2}{2m} \left(\frac{\partial^2}{\partial x^2} + \frac{5}{x} \frac{\partial}{\partial x} + \frac{L^2(\Omega)}{x^2}\right)
\end{equation}
Where $m= \frac{2m_\rho m_\lambda}{m_\rho+m_\lambda}$ is the reduced mass. $L^2(\Omega)$ is the quadratic Casimir operator of the six-dimensional rotational group $O(6)$ and its eigenfunctions are the hyperspherical harmonics. The model Hamiltonian for baryons can be expressed as
\begin{equation}\label{eq:6}
H = \frac{P_x^2}{2m} + V(x)
\end{equation}
The six-dimensional hyperradial Schr\"{o}dinger equation can be written as
\begin{align}\label{eq:7}
\left[\frac{d^2}{dx^2} + \frac{5}{x}\frac{d}{dx} - \frac{\gamma(\gamma + 4)}{x^2} \right] \psi_{\nu\gamma}(x) \nonumber \\
= -2m[E - V(x)]\psi_{\nu\gamma}(x)
\end{align}
Where $\psi_{\nu\gamma}$ is the hyper-radial wave function, labelled by the grand angular quantum number $\gamma$ defined by the number of nodes $\nu$. The potential is assumed to depend only on the hyper radius and hence is a three-body potential since the hyper radius depends only on the coordinates of all three quarks. The hyperCoulomb (hC) plus linear potential, which is given as
\begin{equation}\label{eq:8}
V(x) = \frac{\tau}{x} + \beta x + V_0 + V_{spin}
\end{equation}\\
    Where, $\tau$ = $-$$\frac{2}{3}\alpha_s$ is the hyperCoulomb strength, the values of potential parameter $\beta$ and $V_0$ are fixed to get the ground state masses. $V_{spin}$ is the spin dependent part given as \cite{Garcilazo2007}\\
\begin{equation}\label{eq:9}
V_{spin}(x) =   -\frac{A}{4} \alpha_s \boldsymbol{\lambda}_i \cdot \boldsymbol{\lambda}_j \frac{e^{-x/x_0}}{x {x_0}^2} \sum_{i<j} \frac{\boldsymbol{\sigma}_i \cdot \boldsymbol{\sigma}_j}{6 m_i m_j}
\end{equation}
Here, the parameter $A$ and the regularisation parameter $x_0$ are considered as the hyperfine parameters of the model. The values of parameter $A$ for all SHBs are listed in Table \ref{tab:table3}. The parameter $x_0$ is treated as a hyperfine parameter related to gluon dynamics, independent of the masses of the interacting quarks, as proposed in Ref \cite{majethiya2008a}. As seen in Ref \cite{majethiya2008a}, the values of $A$ vary depending on the quark content, playing a crucial role in determining the mass splittings of singly heavy baryons. There is no well-established procedure to evaluate $x_0$. The hyperfine parameters $A$ and $x_0$ are discussed in detailed, see Ref. \cite{majethiya2008a}. $\boldsymbol{\lambda_{i,j}}$ are the SU(3) colour matrices and $\boldsymbol{\sigma_{i,j}}$ are the spin Pauli matrices, $m_{i,j}$ are the constituent masses of two interacting quarks. The parameter $\alpha_s$ corresponds to the strong running coupling constant, which is given as \cite{Patel2009}
\begin{equation}\label{10}
\alpha_s = \frac{\alpha_s(\mu_0)}{1+(\frac{33-2n_f}{12\pi})\alpha_s(\mu_0)ln(\frac{m_1+m_2+m_3}{\mu_0})}
\end{equation}

      where $\alpha_s(\mu_0 = 1 GeV) \approx 0.6$ is considered in the present study. We factor out the hyperangular part of three-quark wave function, which is given by hyperspherical harmonics. The hyperradial part of the wavefunction is evaluated by solving the Schr\"{o}dinger equation. The hyper-coloumb trial radial wave function, which is given by \cite{Santopinto1998,Ferraris1995}

\begin{align} \label{eq:11}
\psi_{\nu\gamma}=\left[\frac{(\nu - \gamma)! (2g)^6}{(2\nu + 5)(\nu + \gamma + 4)!}\right]^{\frac{1}{2}} (2gx)^\gamma \nonumber\\
\times e^{-gx} L^{2\gamma + 4}_{\nu-\gamma} (2gx)
\end{align}

    Here, $\gamma$ is the hyperangular quantum number and $\nu$ denotes the number of nodes of the spatial three-quark wave function. $L^{2\gamma + 4}_{\nu-\gamma}(2gx)$ is the associated Laguerre polynomial. The wavefunction parameter $g$ and energy eigenvalues are obtained by applying the virial theorem. The masses of ground state singly heavy baryons are calculated by summing the model quark masses (see Table \ref{tab:table2}), kinetic energy and potential energy.
\begin{equation}\label{eq:12}
M_B = m_1 + m_2 + m_3 + \langle H \rangle
\end{equation}

\begin{table}
\centering
    \caption{\label{tab:table2}Quark mass parameters and constants used in the calculations.}
    \begin{tabular}{cc}
    \hline
    Parameter & Value\\
    \hline
    ${m_{u}}$ & 0.33 GeV\\
    ${m_{d}}$ & 0.35 GeV\\
    ${m_{s}}$ & 0.50 GeV\\
    ${m_{c}}$ & 1.55 GeV\\
    ${m_{b}}$ & 4.95 GeV\\
    $\beta$ & 0.14 $GeV^2$\\
    ${V_0}$ for bottom baryons & -0.83 GeV\\
    ${V_0}$ for charmed baryons &-0.79 GeV\\
    $x_0$ & 1.00 $GeV^{-1}$\\
    $\alpha_s(\mu_0 = 1 GeV)$ & 0.6\\
    \hline
    \end{tabular}
\end{table}
\begin{table}
\begin{center}
\centering
    \caption{\label{tab:table3}Hyperfine parameter $A$ in $GeV$ for SHBs.}
    \begin{tabular}{cccccc}
    \hline
    Baryon & quark content & Hyperfine Parameter $A$ & Baryon & quark content & Hyperfine Parameter $A$\\
    \hline
    $\Sigma_c^{++}$	&	uuc	&	14	&	$\Sigma_b^+$	&	uub	&	8	\\
    $\Sigma_c^{+}$	&	udc	&	15	&	$\Sigma_b^0$	&	udb	&	9	\\
    $\Sigma_c^{0}$	&	ddc	&	19	&	$\Sigma_b^-$	&	ddb	&	22	\\
    $\Lambda_c^{+}$	&	udc	&	154	&	$\Lambda_b^0$	&	udb	&	199	\\
    $\Xi_c^{+}$	&	usc	&	72	&	$\Xi_b^0$	&	usb	&	154	\\
    $\Xi_c^{0}$	&	dsc	&	70	&	$\Xi_b^-$	&	dsb	&	150	\\
    $\Omega_c^{0}$	&	ssc	&	30	&	$\Omega_b^-$	&	ssb	&	10	\\
    \hline
    \end{tabular}
\end{center}
\end{table}

\begin{table}[h]
    \caption{{\label{tab:table4} Ground state masses of $J^P=\frac{1}{2}^+$ singly Heavy Baryons in $GeV$}}
    \begin{tabular}{ccccccc}
    \hline
    Baryon	&	Our	&	Experimental \cite{pdg2024}	&	\cite{Farhadi2023}	&	 LQCD \cite{Brown2014}	&	\cite{Pacheco2023}	&	\cite{Wei2017}	\\
    \hline													
    $\Sigma_b^+$	&	5.810	&	$5.81056\pm0.00025$	&	$6.173\pm0.16$	&	 5.856	 &	5.81	&	$5.8134\pm2.8$	\\
    $\Sigma_b^0$	&	5.819	&	-	&	-	&	-	&	-	&	-	\\
    $\Sigma_b^-$	&	5.816	&	$5.81564\pm0.00027$	&	-	&	-	&	-	 &	-	 \\
    $\Lambda_b^0$	&	5.631	&	$5.6196\pm0.00017$	&	$5.534\pm0.15$	&	 5.626	 &	5.615	&	$5.61951\pm0.23$	\\
    $\Xi_b^0$	&	5.791	&	$5.7919\pm0.0005$	&	$5.699\pm0.14$	&	 5.771	&	 5.812	&	$5.7931\pm1.8$	\\
    $\Xi_b^-$	&	5.806	&	$5.7970\pm0.0006$	&	-	&	-	&	-	&		 \\
    $\Omega_b^-$	&	6.013	&	$6.0458\pm0.0008$	&	$6.080\pm0.18$	&	 6.056	 &	6.078	&	$6.048\pm1.9$	\\
    $\Sigma_c^{++}$	&	2.441	&	$2.45397\pm0.00014$	&	$2.404\pm0.08$	&	 2.474	 &	 2.455	&	-	\\
    $\Sigma_c^{+}$	&	2.450	&	$2.45265^{+0.00022}_{-0.00016}$	&	-	&	 -	&	 -	&	-	\\
    $\Sigma_c^{0}$	&	2.453	&	$2.45375\pm0.00014$	&	-	&	-	&	-	 &	-	 \\
    $\Lambda_c^{+}$	&	2.204	&	$2.28645\pm0.0014$	&	$2.475\pm0.09$	&	 2.254	 &	2.281	&	-	\\
    $\Xi_c^{+}$	&	2.463	&	$2.46771\pm0.0023$	&	$2.437\pm0.07$	&	 2.433	&	 2.474	&	-	\\
    $\Xi_c^{0}$	&	2.478	&	$2.47044\pm0.0028$	&	-	&	-	&	-	&	 -	 \\
    $\Omega_c^{0}$	&	2.638	&	$2.6952\pm0.017$	&	$2.592\pm0.09$	&	 2.679	 &	2.733	&	-	\\
    \hline
    \end{tabular}
\end{table}

\begin{table}[h]
    \caption{{\label{tab:table5} Ground state masses of $J^P=\frac{3}{2}^+$ singly Heavy Baryons in $GeV$}}
    \begin{tabular}{ccccccc}
    \hline
    Baryon	&	Our	&	Experimental \cite{pdg2024}	&	LQCD \cite{Brown2014}	 &	 \cite{Pacheco2023}	&	\cite{Wei2017}	&	\cite{Weng2024}	\\
    \hline												
    $\Sigma_b^{+*}$	&	5.823	&	$5.83031\pm0.00027$	&	5.877	&	 $5.832\pm0.003$	&	$5.8336\pm2.4$	&	5.838	\\
    $\Sigma_b^{0*}$	&	5.833	&	-	&	-	&	-	&	-	&	-	\\
    $\Sigma_b^{-*}$	&	5.848	&	$5.83474\pm0.00030$	&	-	&	-	&	-	 &	-	 \\
    $\Xi_b^{0*}$	&	5.985	&	$5.9523\pm0.0006$	&	5.960	&	-	&	 $5.9521\pm3.3$	 &	5.964	\\
    $\Xi_b^{-*}$	&	5.992	&	$5.9557\pm0.0005$	&	-	&	-	&	-	 &	-	 \\
    $\Omega_b^{-*}$	&	6.024	&	-	&	6.085	&	$6.100\pm0.003$	&	 $6.0694\pm6.9$	&	-	\\
    $\Sigma_c^{++*}$	&	2.479	&	$2.51841\pm0.00022$	&	2.551	&	 $2.521\pm0.002$	&	-	&	2.518	\\
    $\Sigma_c^{+*}$	&	2.490	&	$2.5174^{+0.007}_{-0.005}$	&	-	&	-	 &	-	 &	-	\\
    $\Sigma_c^{0*}$	&	2.503	&	$2.51848\pm0.00021$	&	-	&	-	&	-	 &	-	 \\
    $\Lambda_c^{+*}$	& 2.613   &	 $2.8561^{+0.002}_{-0.0017}\pm0.0005^{+0.0011}_{-0.0056}$	&	-	&	-	 &	-	 &	2.871	\\
    $\Xi_c^{+*}$	&	2.627	&	$2.6451\pm0.0003$	&	2.648	&	-	&	 -	&	 2.649	\\
    $\Xi_c^{0*}$	&	2.634	&	$2.6457\pm0.00025$	&	-	&	-	&	-	 &	-	 \\
    $\Omega_c^{0*}$	&	2.696	&	$2.7659\pm0.0002$	&	2.755	&	 $2.799\pm0.002$	&	-	&	2.768	\\
    \hline
\end{tabular}
\end{table}
\section{Magnetic moments and radiative decay}\label{sec:2}
Understanding electromagnetic properties and radiative transitions of heavy baryons are the keys to heavy flavour dynamics. The observation of the magnetic moments and transition decay widths provide the testing hypothesis for various theoretical models on the hadron structure. Experimentally, $\Omega_c^{+*} \rightarrow \Omega_c^+\gamma$ \cite{Aubert2006,Solovieva2009}, $\Xi_c^{'+} \rightarrow \Xi_c^{+}\gamma$ and $\Xi_c^{'0} \rightarrow \Xi_c^0\gamma$ \cite{Jessop1999,Aubert2018,Yelton2016} radiative decay processes have been observed by BABAR and Belle Collaborations. Theoretically, the magnetic moments and radiative decays are studied within different approaches such as Effective mass scheme \cite{Hazra2021,Dhir2009}, Chiral perturbation theory \cite{Wang2019}, MIT bag model \cite{Hackman1978}, Lattice QCD \cite{Can2015,Bahtiyar2017}, Light cone QCD sum rule \cite{Aliev2009}. In this work, we have computed the magnetic moments, transition magnetic moments and radiative $M1$ decay widths for SHBs. The magnetic moments of the baryons are computed in terms of the spin-flavour wave function of the constituent quarks as
\begin{equation}\label{eq:13}
\mu_B = \Sigma_i \langle \phi_{sf}|\mu_i \boldsymbol{\sigma_i}|\phi_{sf}\rangle
\end{equation}
where
\begin{equation}\label{eq:14}
\mu_i = \frac{e_i}{2m_i}
\end{equation}
Here, $e_i$ and $\boldsymbol{\sigma_i}$ are the charge and the spin of the quark. $|\phi_{sf}\rangle$ represents the spin-flavour wave function of the respective baryonic state and $m_i$ is the mass of the $i^{th}$ quark of the baryon. As the magnetic moment of a baryon is computed in terms of its constituting quark content, we must consider the bound state effect in the calculations of the magnetic moments of the bound quarks. Such effects are incorporated by defining an effective mass for the quarks ($m^{eff}_i$) constituting the baryonic state. To account the bound state effect, we replace the quark mass parameter $m_i$ of Eqn. (\ref{eq:14}) by defining an effective mass for the bound quarks $m^{eff}_i$ as
\begin{equation}\label{eq:15}
m_i^{eff} = m_i \left(1+\frac{\langle H \rangle}{\sum_i m_i}\right)
\end{equation}
such that $M_B = \sum_{i=1}^3 m_i^{eff} $ where $\langle H \rangle$ = E + $\langle V(x) \rangle$. The computations are repeated for different choices of flavour combinations of $qqQ$ ($q = u, d, s$ and $Q = b, c$).
The transition magnetic moment for $\frac{3}{2}^+ \rightarrow \frac{1}{2}^+$ can be expressed as \cite{Thakkar2011}
\begin{equation}\label{eq:16}
\mu_{\frac{3}{2}^+ \rightarrow \frac{1}{2}^+} = \sum_i \left\langle \phi_{sf}^{\frac{3}{2}^+} | \mu_i\boldsymbol{\sigma_i}| \phi_{sf}^{\frac{1}{2}^+} \right\rangle
\end{equation}
$\langle \phi_{sf} |$ represent the spin-flavour wave function of the quark composition for the respective baryons. To compute the transition magnetic moment ($\mu_{\frac{3}{2}^+ \rightarrow \frac{1}{2}^+}$), we take the geometric mean of effective quark masses of the constituent quarks of initial and final state baryons,
\begin{equation}\label{eq:17}
m_{i{B^* \rightarrow B}}^{eff} = \sqrt{m_{i{B^*}}^{eff}m_{iB}^{eff}}
\end{equation}
Here, $m_{i{B^*}}^{eff}$ and $m_{iB}^{eff}$ are the effective masses of the quarks constituting the baryonic states $B^*$ and $B$, respectively. The expressions and the obtained transition magnetic moments of singly heavy baryons are listed in Table \ref{tab:table9}.
In our calculation od radiative decay width, we ignore $E2$ amplitudes because of the spherical symmetry of the S-wave baryon spatial wave function and radiative $M1$ decay width can be expressed in terms of the radiative transition magnetic moment and the photon momentum ($k$) as \cite{Wagner2000,{Bernotas2013}}
\begin{equation}\label{eq:18}
 \Gamma = \frac{\alpha k^3}{M_P^2} \frac{2}{2J+1}\frac{M_B}{M_{B^*}} \mu^2(B^* \rightarrow B)
\end{equation}
where, $\mu^2(B^* \rightarrow B\gamma)$ is square of the transition magnetic moment, $\alpha=\frac{1}{137}$ and $M_P$  = 0.938 $GeV$ is mass of proton. $J$ and $M_{B^*}$ are the total angular momentum and mass of the decaying baryon and $M_B$ is the baryon mass of the final state. $k$ is the photon momentum in the center-of-mass system of decaying baryon given as
\begin{equation}\label{eq:19}
k = \frac{M^2_{B^*} - M^2_B}{2M_B}
\end{equation}
The calculated radiative $M1$ decay widths of SHBs are listed in Table \ref{tab:table10}.

\begin{table}[h]
    \caption{\label{tab:table6}Magnetic moments of $J^P=\frac{1}{2}^+$ singly bottom Baryons in $\mu_N$}
    \centering
    \begin{tabular}{cccccccc}
    \hline
    Baryon	&	Our	&	EMS \cite{Hazra2021}	&	LCQSR \cite{Ozdem2024}	&	 PM \cite{Barik1983}	&	BM \cite{Bernotas13}	&	RTQM \cite{Faessler2006}	 &	 BM \cite{Simonis2018}	\\
    \hline																
    $\Sigma_b^+$	&	2.460	&	$2.1989\pm0.0021$	&	$2.02\pm0.19$	&	 2.575	 &	1.622	&	2.07	&	2.250		\\
    $\Sigma_b^0$	&	0.683	&	$0.5653\pm0.0011$	&	$0.53\pm0.06$	&	 0.659	 &	0.422	&	0.53	&	0.603		\\
    $\Sigma_b^-$	&	-1.137	&	$-1.0684\pm0.0011$	&	$-1.01\pm0.9$	&	 -1.256	&	-0.778	&	-1.01	&	-1.150		\\
    $\Lambda_b^0$	&	-0.063	&	$-0.06202\pm0.00001$	&	$-0.29\pm0.03$	 &	-	 &	-0.066	&	-0.06	&	-0.060		\\
    $\Xi_b^0$	&	0.866	&	$-0.06202\pm0.00001$	&	$-0.33\pm0.03$	&	 -	&	 -0.100	&	-0.06	&	-0.060	\\
    $\Xi_b^-$	&	-0.990	&	$-0.06202\pm0.00001$	&	$-0.25\pm0.03$	&	 -	&	 -0.063	&	-0.06	&	-0.055		\\
    $\Omega_b^-$	&	-0.804	&	$-0.7454\pm0.0024$	&	$-0.87\pm0.07$	&	 -0.714	&	-0.545	&	-0.82	&	-0.806		\\
    \hline
    \end{tabular}
\end{table}

\begin{table}[h]
    \caption{\label{tab:table7}Magnetic moments of $J^P=\frac{1}{2}^+$ singly charmed Baryons in $\mu_N$}
    \centering
    \begin{tabular}{cccccccc}
    \hline
    Baryon	&	Our	&	EMS \cite{Hazra2021}	&	BM \cite{Simonis2018}	&	 ChPT \cite{Shi2019}	&	hCQM \cite{Gandhi2018}	&	CQSM \cite{Yang2018}	 &	LCQSR \cite{Ozdem2024} \\
    \hline																
    $\Sigma_c^{++}$	&	2.166	&	$2.0932\pm0.00032$	&	2.280	&	2.00	 &	 1.831	&	$2.15\pm0.10$	&	$2.02\pm0.18$		\\
    $\Sigma_c^{+}$	&	1.216	&	$0.42707\pm0.0017$	&	0.487	&	0.46	 &	 0.380	&	$0.46\pm0.03$	&	$0.50\pm0.05$		\\
    $\Sigma_c^{0}$	&	0.135	&	$-1.2392\pm0.0017$	&	-1.310	&	-1.08	 &	 -1.091	&	$-1.24\pm0.05$	&	$-1.01\pm0.09$		\\
    $\Lambda_c^{+}$	&	0.408	&	$0.3801\pm0.0008$	&	0.335	&	0.24	 &	 0.421	&	-	&	$0.46\pm0.09$		\\
    $\Xi_c^{+}$	&	0.688	&	$0.3801\pm0.0008$	&	0.142	&	0.24	&	 -	&	 -	&	$0.33\pm0.05$		\\
    $\Xi_c^{0}$	&	0.363	&	$0.3801\pm0.0008$	&	0.346	&	0.19	&	 -	&	 -	&	$0.41\pm0.05$		\\
    $\Omega_c^{0}$	&	-0.936	&	$-0.9057\pm0.0021$	&	-0.950	&	-0.74	 &	 -1.179	&	$-0.85\pm005$	&	$-0.73\pm0.08$		\\
    \hline
    \end{tabular}
\end{table}

\begin{table}
    \caption{\label{tab:table8}Magnetic moments of $J^P=\frac{3}{2}^+$ singly Heavy Baryons in $\mu_N$}
    \tiny
    \centering
    \begin{tabular}{cccccccc}
    \hline
    Baryon	&	Our	&	EMS \cite{Hazra2021}	&	LCQSR \cite{Ozdem2024}	&	 MITBM \cite{Bernotas13}	&	$\chi$QSM \cite{Yang2018}	&	BM \cite{Simonis2018}	&	 QCDSR \cite{Aliev9}		\\
    \hline																
    $\Sigma_b^{+*}$	&	3.591	&	$3.1637\pm0.0031$	&	$3.20\pm0.26$	&	 3.56	 &	-	&	3.46	&	2.52		\\
    $\Sigma_b^{0*}$	&	0.907	&	$0.7444\pm0.0016$	&	$0.93\pm0.10$	&	 0.87	 &	-	&	0.82	&	0.50	\\
    $\Sigma_b^{-*}$	&	-1.787	&	$-1.6748\pm0.0015$	&	$-1.54\pm0.12$	&	 -1.92	 &	-	&	-1.82	&	-1.50		\\
    $\Xi_b^{0*}$	&	1.204	&	$1.0151\pm0.0019$	&	$0.48\pm0.05$	&	 1.19	 &	-	&	1.03	&	0.50	\\
    $\Xi_b^{-*}$	&	-1.531	&	$-1.4379\pm0.0015$	&	$-0.57\pm0.05$	&	 -1.6	 &	-	&	-1.55	&	-1.42		\\
    $\Omega_b^{-*}$	&	-1.297	&	$-1.1985\pm0.0030$	&	$-0.61\pm0.05$	&	 -1.28	 &	-	&	-1.31	&	-1.40		\\
    $\Sigma_c^{++*}$	&	3.738	&	$3.5730\pm0.004$	&	$3.40\pm0.34$	 &	 4.11	&	$3.22\pm0.15$	&	3.98	&	4.81		\\
    $\Sigma_c^{+*}$	&	1.258	&	$1.1763\pm0.002$	&	$0.90\pm0.10$	&	 1.32	 &	$0.68\pm0.04$	&	1.25	&	2.00		\\
    $\Sigma_c^{0*}$	&	-1.243	&	$-1.2198\pm0.0018$	&	$-1.44\pm0.19$	&	 -1.47	 &	$-1.50\pm0.07$	&	-1.49	&	-0.81		\\
    $\Xi_c^{+*}$	&	1.617	&	$1.4426\pm0.0022$	&	$0.80\pm0.08$	&	 1.64	 &	$0.75\pm0.04$	&	1.47	&	1.68		\\
    $\Xi_c^{0*}$	&	-1.016	&	$-0.9866\pm0.0017$	&	$-0.51\pm0.05$	&	 -1.15	 &	$-1.50\pm0.07$	&	-1.20	&	-0.68		\\
    $\Omega_c^{0*}$	&	-0.801	&	$-0.7512\pm0.0029$	&	$-0.70\pm0.05$	&	 -0.86	 &	$-1.50\pm0.07$	&	-0.936	&	-0.62		\\
    \hline
    \end{tabular}
\end{table}

\begin{table}
    \caption{\label{tab:table9}Transition magnetic moments of singly Heavy Baryons in $\mu_N$}
    \tiny
    \begin{tabular}{ccccccccc}
    \hline
    Transition	&	$\mu$	&	Our	&	\cite{Bernotas2013}	&	NRQM \cite{Bernotas2013}	&	ChPT \cite{Sharma2010}	&	CQSM \cite{Yang2020}	 &	 EMS \cite{Hazra2021}	 & BM	 \cite{Simonis2018}	\\
    \hline																
    $\Sigma_b^{+*}\rightarrow\Sigma_b^+$	&	 $\frac{2\sqrt{2}}{3}(\mu_u-\mu_b)$	&	 1.781	&	1.193	&	1.81	&	 -	&	$-1.52\pm0.07$	&	$1.5889\pm0.001$	 &	 1.690	 \\
    $\Sigma_b^{0*}\rightarrow\Sigma_b^0$	&	 $\frac{\sqrt{2}}{3}(\mu_u+\mu_d-2\mu_b)$	&	0.514	&	0.345	&	0.49	 &	 -	&	$-0.33\pm0.02$	&	 $0.4411\pm0.0006$	&	 0.464	\\
    $\Sigma_b^{-*}\rightarrow\Sigma_b^-$	&	 $\frac{2\sqrt{2}}{3}(\mu_d-\mu_b)$	&	 -0.758 &	-0.504	&	-0.82	&	 -	&	$0.87\pm0.03$	&	$-0.7067\pm0.0005$	 &	 -0.760	 \\
    $\Xi_b^{0*}\rightarrow\Xi_b^0$	&	$\sqrt{\frac{2}{3}}(\mu_u-\mu_s)$	&	 2.046	 &	1.321	&	2.03	&	-	&	$1.69\pm0.08$	&	 $1.9887\pm0.0015$	&	 1.830	 \\
    $\Xi_b^{-*}\rightarrow\Xi_b^-$	&	$\sqrt{\frac{2}{3}}(\mu_d-\mu_s)$	&	 -0.227	&	-0.139	&	-0.26	&	-	&	$-0.29\pm0.04$	&	 $-0.2570\pm0.001$	 &	 -0.182	 \\
    $\Omega_b^{-*}\rightarrow\Omega_b^-$	&	 $\frac{\sqrt{2}}{3}(\mu_s-\mu_c)$	&	 -0.524	&	-0.339	&	-0.52	&	 -	&	$0.60\pm0.04$	&	$-0.4802\pm0.0011$	 &	 -0.523	\\
    $\Sigma_c^{++*}\rightarrow\Sigma_c^{++}$	&	 $\frac{2\sqrt{2}}{3}(\mu_u-\mu_c)$	&	1.264	&	0.905	&	1.39	&	 -1.37	 &	$1.52\pm0.07$	&	 $1.1786\pm0.0015$	&	 1.340	\\
    $\Sigma_c^{+*}\rightarrow\Sigma_c^{+}$	&	 $\frac{\sqrt{2}}{3}(\mu_u+\mu_d-2\mu_c)$	&	0.083	&	-0.062	&	0.07	 &	 -0.003	&	$0.33\pm0.02$	&	 $0.025\pm0.0009$	 &	0.102	\\
    $\Sigma_c^{0*}\rightarrow\Sigma_c^{0}$	&	 $\frac{2\sqrt{2}}{3}(\mu_d-\mu_c)$	&	 -1.110	&	-1.030	&	-1.24	&	 1.48	&	$-0.87\pm0.03$	&	 $-1.1286\pm0.0009$	 &	 -1.140	\\
    $\Xi_c^{+*}\rightarrow\Xi_c^{+}$	&	$\sqrt{\frac{2}{3}}(\mu_u-\mu_s)$	 &	 1.913	&	1.497	&	2.03	&	2.08	&	$1.69\pm0.08$	&	 $1.9797\pm0.0015$	&	 1.860	 \\
    $\Xi_c^{0*}\rightarrow\Xi_c^{0}$	&	$\sqrt{\frac{2}{3}}(\mu_d-\mu_s)$	 &	 -0.227	&	-0.224	&	-0.33	&	-0.50	&	$-0.29\pm0.04$	&	 $-0.255\pm0.001$	&	 -0.249	\\
    $\Omega_c^{0*}\rightarrow\Omega_c^{0}$	&	 $\frac{2\sqrt{2}}{3}(\mu_s-\mu_c)$	&	 -0.927	&	-0.839	&	-0.94	&	 0.96	&	$-0.60\pm0.04$	&	 $-0.9004\pm0.0011$	 &	 -0.892	\\
    \hline
    \end{tabular}
\end{table}


\begin{sidewaystable*}
    \caption{\label{tab:table10}Radiative M1 decay widths of singly Heavy Baryons in $keV$}
    \small
    \begin{tabular}{cccccccccccc}
    \hline
    Transition	&	Our	&	\cite{Bernotas2013}	&	CQSM \cite{Yang2020}	&	 EMS \cite{Hazra2021}	&	BM \cite{Simonis2018}	&	LCQSR \cite{Aliev2009}	&	 \cite{Peng2024}	 &	 \cite{Wang2017}	&	 \cite{Pacheco2023}	&	 \cite{Wang2019}	&	\cite{Farhadi2023}			 \\
    \hline																									
    $\Sigma_b^{+*}\rightarrow\Sigma_b^+\gamma$	&	0.026	&	0.054	&	 0.0022	&	 $0.080\pm0.004$	&	0.11	&	$0.46\pm0.22$	&	0.1	&	0.25	 &	0.1	&	 0.05	 &	 $0.116\pm0.017$			\\
    $\Sigma_b^{0*}\rightarrow\Sigma_b^0\gamma$	&	0.003	&	0.005	&	 0.0001	&	 $0.0059\pm0.0002$	&	0.008	&	$0.028\pm0.016$	&	0.0	&	 0.02	&	0.0	&	 0.003	&	 $0.0082\pm0.001$			\\
    $\Sigma_b^{-*}\rightarrow\Sigma_b^-\gamma$	&	0.077	&	0.010	&	 0.001	&	 $0.0144\pm0.0009$	&	0.019	&	$0.11\pm0.06$	&	0.0	&	 0.06	&	0.0	&	 0.013	&	 $0.0221\pm0.008$			\\
    $\Xi_b^{0*}\rightarrow\Xi_b^0\gamma$	&	118.357	&	24.7	&	 $1.2\pm0.1$	 &	$65.0\pm0.9$	&	55.3	&	$135\pm65$	&	40.8	&	 104	&	-	&	 17.2	 &	 $132.38\pm2.005$		\\
    $\Xi_b^{-*}\rightarrow\Xi_b^-\gamma$	&	1.264	&	0.278	&	 $0.03\pm0.01$	 &	$1.044\pm0.015$	&	0.536	&	$1.50\pm0.75$	&	0.8	 &	0.00	&	-	&	 1.40	 &	 $1.43\pm0.039$				\\
    $\Omega_b^{-*}\rightarrow\Omega_b^-\gamma$	&	0.001	&	0.006	&	 $0.004\pm0.001$	&	$0.056\pm0.008$	&	0.009	&	-	&	0.0	&	0.10	&	 0.00	&	 0.031	 &	 $0.00071\pm0.0001$			\\
    $\Sigma_c^{++*}\rightarrow\Sigma_c^{++}\gamma$	&	0.355	&	0.826	&	 $0.36\pm0.03$	&	$1.483\pm0.018$	&	1.96	&	$2.65\pm1.60$	&	1.7	 &	 3.94	&	2.1	 &	 1.20	&	$6.29\pm0.22$			\\
    $\Sigma_c^{+*}\rightarrow\Sigma_c^{+}\gamma$	&	0.002	&	0.004	&	 $0.02\pm0.003$	&	$(6.7\pm0.9)\times10^{-4}$	&	0.011	&	 $0.40\pm0.16$	 &	 0.0	&	 0.004	 &	 0.0	&	0.04	&	$0.39\pm0.037$			 \\
    $\Sigma_c^{0*}\rightarrow\Sigma_c^{0}\gamma$	&	0.633	&	1.08	&	 $0.08\pm0.01$	&	$1.378\pm0.016$	&	1.41	&	$0.08\pm0.03$	&	1.3	 &	 3.43	&	1.8	 &	 0.49	&	$1.53\pm0.051$			\\
    $\Xi_c^{+*}\rightarrow\Xi_c^{+}\gamma$	&	85.719	&	44.3	&	 $8.66\pm0.81$	 &	$81.9\pm0.5$	&	81.6	&	$52\pm25$	&	52.7	 &	139	&	-	&	 0.07	 &	 $49.45\pm1.61$			\\
    $\Xi_c^{0*}\rightarrow\Xi_c^{0}\gamma$	&	0.697	&	0.908	&	 $0.25\pm0.06$	 &	$1.322\pm0.014$	&	0.745	&	$0.66\pm0.32$	&	1.1	 &	0.0	&	-	&	 0.42	 &	 $0.63\pm0.032$			\\
    $\Omega_c^{0*}\rightarrow\Omega_c^{0}\gamma$	&	0.639	&	1.07	&	 $0.06\pm0.01$	&	$1.14\pm0.13$	&	1.13	&	-	&	0.9	&	0.89	 &	 1.0	&	0.32	 &	 $1.11\pm0.014$			\\
    \hline
    \end{tabular}
\end{sidewaystable*}

\section{Exclusive semileptonic decay $b \rightarrow c l\bar{\nu_l}$}\label{sec:3}
The semileptonic decays of singly heavy baryons have been studied within different theoretical frameworks such as Bethe Salpeter approach \cite{Ivanov1999}, Quark-diquark model \cite{Nejad2020}, Non-relativistic quark model \cite{Cheng1996}, Relativistic quark model \cite{Ebert2006}, Effective field theory \cite{Rajeev2019}, Bethe-Salpeter approach \cite{Rusetsky1997}. Different forms of IWF are studied in Ref. \cite{{Thakkar2024},{Jenkins1993},{Sadzikowski1993},{Guo1993},{Becirevic2020}}. Especially, the IWF and semileptonic transition of $\Lambda_b^0$ baryon has been studied widely, see Ref.  \cite{{Faustov2016},{Thakkar2020},{Guo1996},{Azizi2018},{Woloshyn2013}}. The IWF is crucial for extracting CKM matrix elements (e.g., $|V_{cb}|$). In HQET, when $Q$ ($Q = b, c$) is sufficiently heavy ($m_Q \rightarrow \infty$), the heavy quark acts like a static source of gluons in its rest frame and the internal dynamics of heavy baryons becomes independent of $m_Q$. The spin of the heavy quark decouples from the light quark and gluon degrees of freedom. This flavour and spin symmetry provide several model independent relations for the heavy-to-heavy baryonic form factors. In the approximation of infinite heavy quark masses, the six form factors $F_i$, $G_i$  $(i = 1, 2, 3)$ can be expressed in terms of a single universal Isgur-Wise function $\xi(\omega)$.

\begin{figure}[h]
    \centering
    \begin{tikzpicture}
        \begin{feynman}
            \vertex (a) {$qqb$};
            \vertex[right=2cm of a] (b);
            \vertex[right=2cm of b] (c) {$qqc$};
            \vertex[above right=1.5cm of b] (w);
            \vertex[above left=1.5cm of w] (nubar) {$\bar{\nu}_\ell$};
            \vertex[above right=1.5cm of w] (l) {$\ell^-$};
            \diagram* {
                (a) -- [fermion] (b) -- [fermion] (c),
                (b) -- [boson, edge label={$W^-$}] (w),
                (w) -- [fermion] (l),
                (w) -- [anti fermion] (nubar)
                };
        \end{feynman}
    \end{tikzpicture}
    \caption{Feynman diagram for semileptonic decay of singly heavy baryons, $qqb \rightarrow qqc \ell^- \bar{\nu}_\ell$}
    \label{fig:1}
\end{figure}
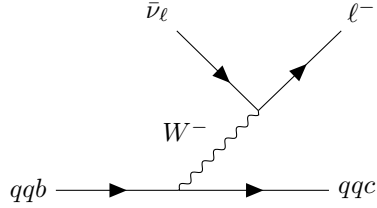

\begin{equation}\label{eq:20}
F_1(q^2) = G_1(q^2) = \xi(\omega)
\end{equation}
\begin{equation}\label{eq:21}
F_2=F_3=G_2=G_3=0
\end{equation}
where, $\omega=\nu\cdot\nu'$ is the velocity transfer between the initial $\nu$ and final $\nu'$ heavy baryons. It is related to the squared four-momentum transfer between the heavy baryons, $q^2$, by the equation
\begin{equation}\label{eq:22}
\omega=\frac{m^2_{b}+m^2_{c}-q^2}{2m_{b}m_{c}}
\end{equation}
Here, we have calculated the IWF at zero recoil point ($\xi(\omega)|_{\omega=1}=1$), using Taylor's series expansion as
\begin{equation}\label{eq:23}
\xi(\omega)=1-\rho^2 (\omega-1)+c(\omega-1)^2+...
\end{equation}
where $\rho^2$ is the magnitude of the slope and $c$ is the curvature (convexity parameter) of IWF ($\xi(\omega)$). $\rho^2$ and c can be written as
\begin{equation}\label{eq:24}
\rho^2= - \frac{d\xi(\omega)}{d\omega}|_{\omega=1}
\end{equation}
\begin{equation}\label{eq:25}
c=\frac{d^2\xi(\omega)}{d\omega^2}|_{\omega=1}
\end{equation}
The higher terms in the IWF may be negligible since the slope and curvature are the most dominant parameters. In hCQM, the IWF can be written in terms of the overlap integral of the initial and final baryon wave function as \cite{Hassanabadi2014}

\begin{equation}\label{eq:26}
\xi(\omega)=16\, \pi^2\int_{0}^{\infty} |\psi_{\nu\gamma}(x)|^2\, cos(px)\, x^5 \,dx
\end{equation}
Because we have investigated the Isgur-Wise function near the zero recoil point ($\omega=1$), only $|\psi(x)|^2$ is considered instead of the overlap integral, where the four velocities of the baryons before and after transitions are identical. Here, $\nu=\gamma=0$ as we study the IWF for ground state heavy baryons. If $\cos(px)$ is expanded, then
\begin{equation}\label{eq:27}
cos(p x)= 1-\frac{p^2x^2}{2!}+\frac{p^4x^4}{4!}+...
\end{equation}
where $p^2$ is the square of virtual momentum transfer which can be written as $p^2 = 2 m^2 (\omega-1)$. We get the slope and curvature of the IWF by substituting Eqn. (\ref{eq:27}) into Eqn. (\ref{eq:26}) and then comparing Eqn. (\ref{eq:26}) with Eqn. (\ref{eq:23}) as
\begin{equation}\label{eq:28}
\rho^2=16 \pi^2 m^2\int_{0}^{\infty} |\psi_{\nu\gamma}(x)|^2  x^7 dx
\end{equation}
\begin{equation}\label{eq:29}
c=\frac{8}{3} \pi^2 m^4\int_{0}^{\infty} |\psi_{\nu\gamma}(x)|^2  x^9 dx
\end{equation}
At zero recoil point ($\omega=1$), the IWF is normalised, $\xi(\omega)=1$. The semileptonic transition of heavy baryons can be predicted using the obtained IWF. The differential decay width for the semileptonic transition of heavy baryons can be written as \cite{Guo1996}
\begin{eqnarray}\label{eq:30}
\frac{d\Gamma}{d\omega}=\frac{2}{3} m^4_{B_c} m_{B_b} A \xi^2(\omega) \sqrt{\omega^2-1}\nonumber\\
\times \left[3 \omega (\eta+\eta^{-1})-2-4\omega^2\right]
\end{eqnarray}
where, $A=\frac{G^2_{F}}{(2\pi)^3}$ $|V_{cb}|^2$ $Br(B_c \rightarrow ab)$. $G_{F}$ is the Fermi coupling constant and $|V_{cb}|$ is the Kobayashi-Maskawa matrix element.
$\eta=m_{B_b}/m_{B_c}$ and $Br(B_c \rightarrow ab)$ is branching ratio. $m_{B_b}$ and $m_{B_c}$ are the masses of bottom baryon and charmed baryon, respectively. The total decay width is obtained by the integration of Eqn. (\ref{eq:30}) as
\begin{equation}\label{eq:31}
\Gamma=\int_{1}^{\omega_{max}}\frac{d\Gamma}{d\omega} d\omega
\end{equation}
where $\omega_{max}$ is the maximal recoil ($q^2=0$) and it can be written as
\begin{equation}\label{eq:32}
\omega_{max}=\frac{m^2_{B_b}+m^2_{B_c}}{2\,m_{B_b}m_{B_c}}
\end{equation}


\begin{figure}
\centering
\includegraphics[scale=0.33]{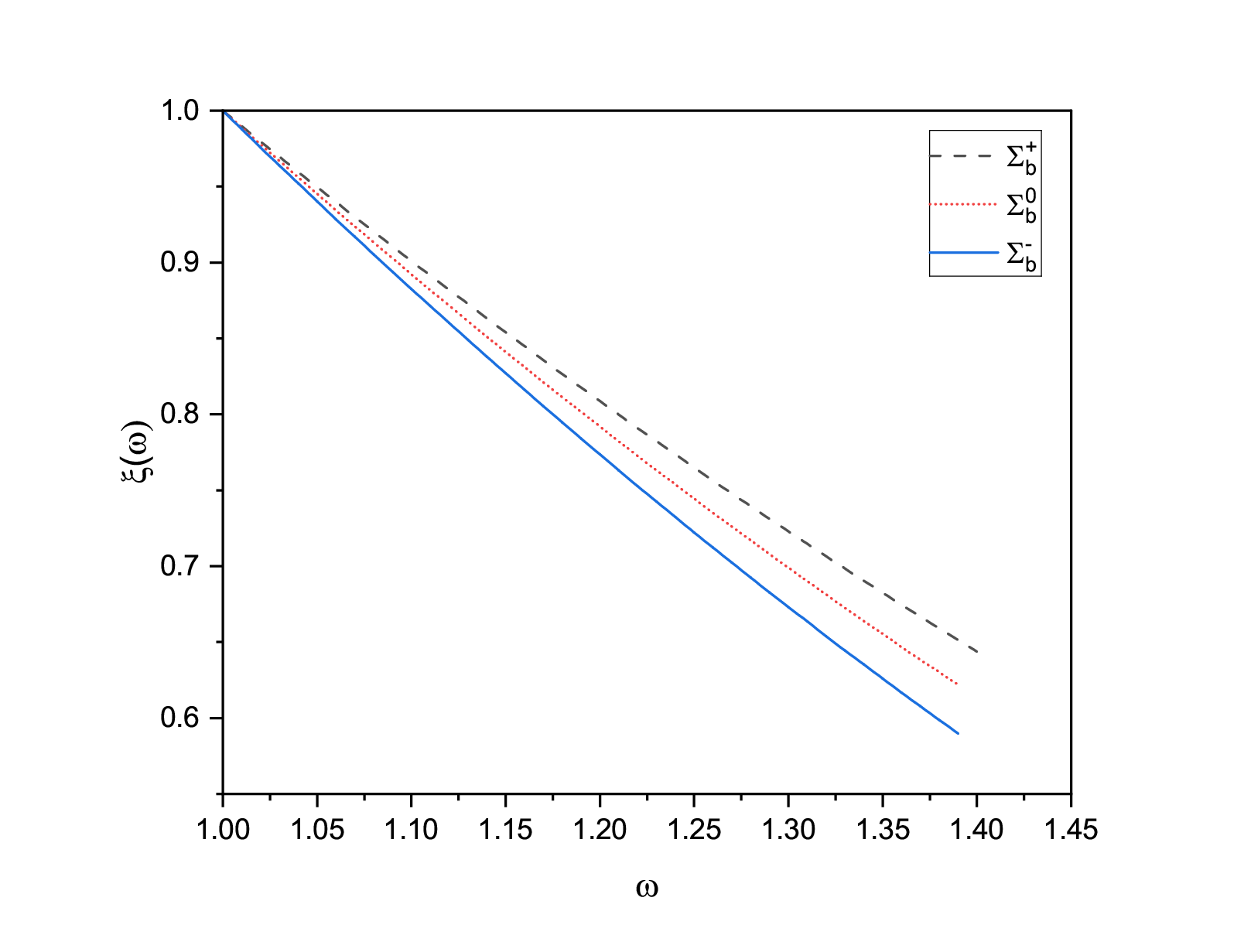}
\caption{\label{fig:2}The Isgur-Wise function $\xi(\omega)$ for $\Sigma_b^{+(0,-)} \rightarrow \Sigma_c^{++(+,0)} \ell\bar{\nu}$ transition}
\end{figure}

\begin{figure}
\centering
\includegraphics[scale=0.33]{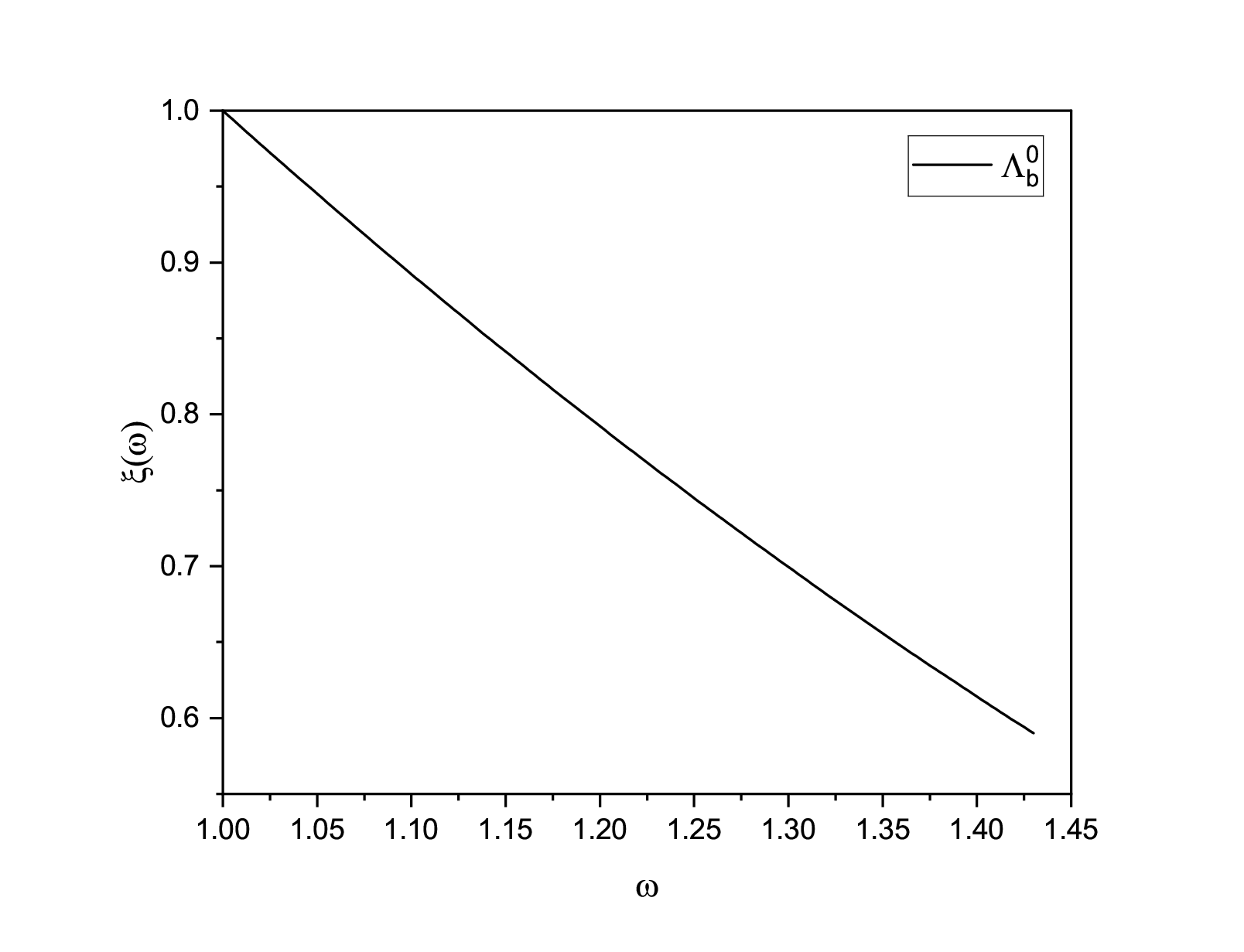}
\caption{\label{fig:3}The Isgur-Wise function $\xi(\omega)$ for $\Lambda_{b}^0 \rightarrow \Lambda_{c}^+ \ell\bar{\nu}$ transition}
\end{figure}
\begin{figure}
\centering
\includegraphics[scale=0.33]{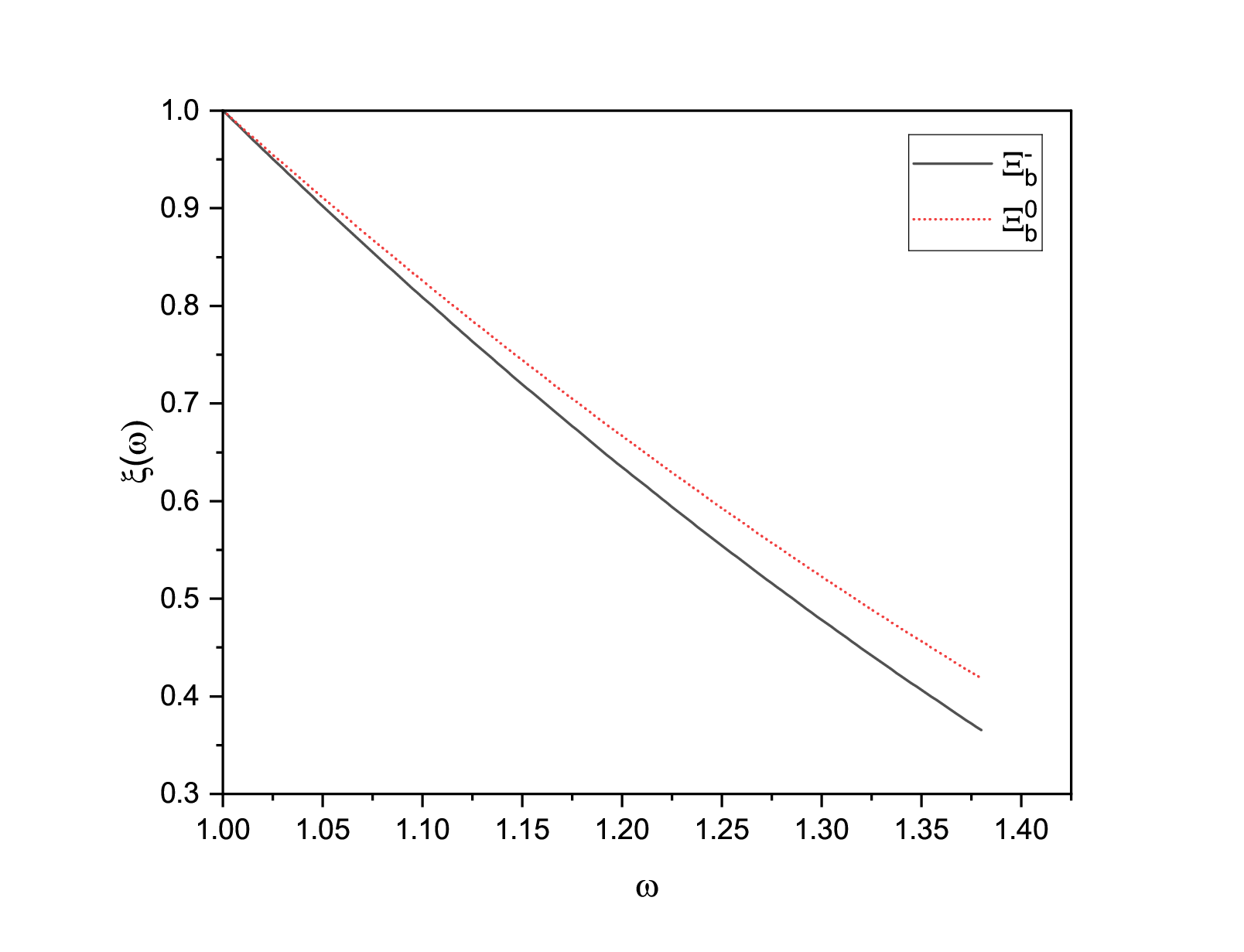}
\caption{\label{fig:4}The Isgur-Wise function $\xi(\omega)$ for $\Xi_{b}^{0(-)} \rightarrow \Xi_{c}^{+(0)} \ell\bar{\nu}$ transition}
\end{figure}

\begin{figure}
\centering
\includegraphics[scale=0.33]{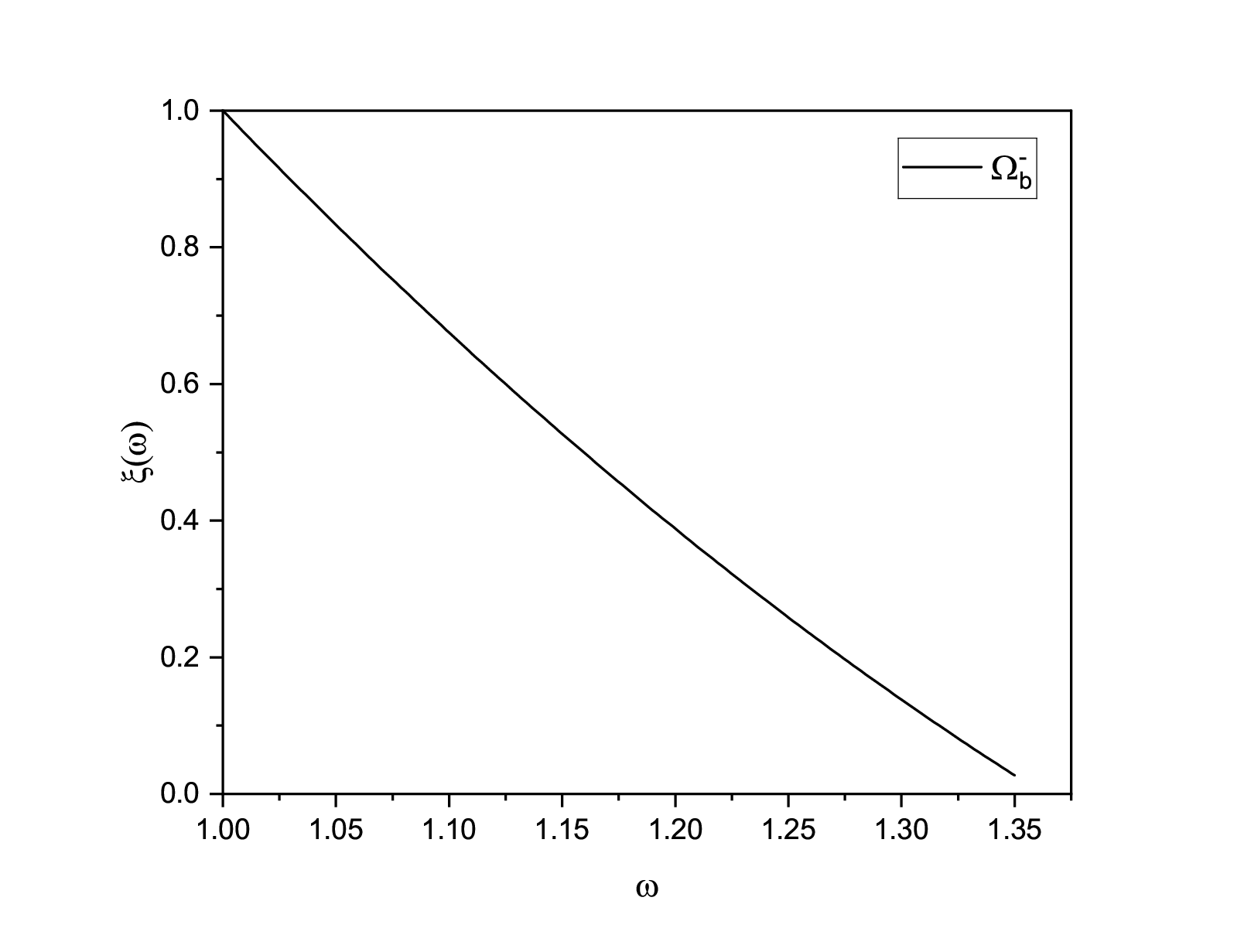}
\caption{\label{fig:5}The Isgur-Wise function $\xi(\omega)$ for $\Omega_{b}^- \rightarrow \Omega_{c}^0 \ell\bar{\nu}$ transition}
\end{figure}

\begin{figure}
\centering
\includegraphics[scale=0.33]{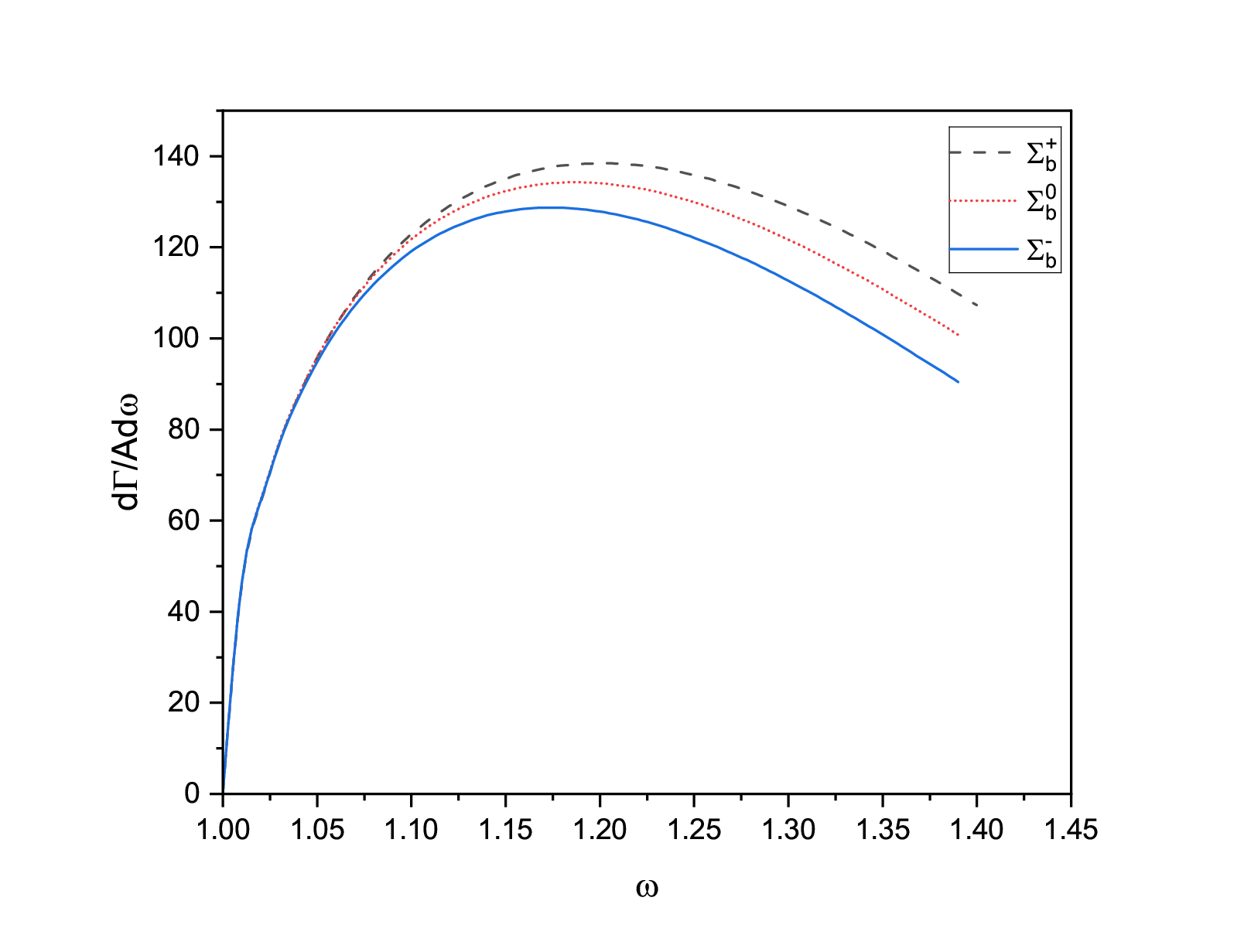}
\caption{\label{fig:6}Differential decay rate for $\Sigma_b^{++(+,0)} \rightarrow \Sigma_c^{++(+,0)} \ell\bar{\nu}$ transition}
\end{figure}

\begin{figure}
\centering
\includegraphics[scale=0.33]{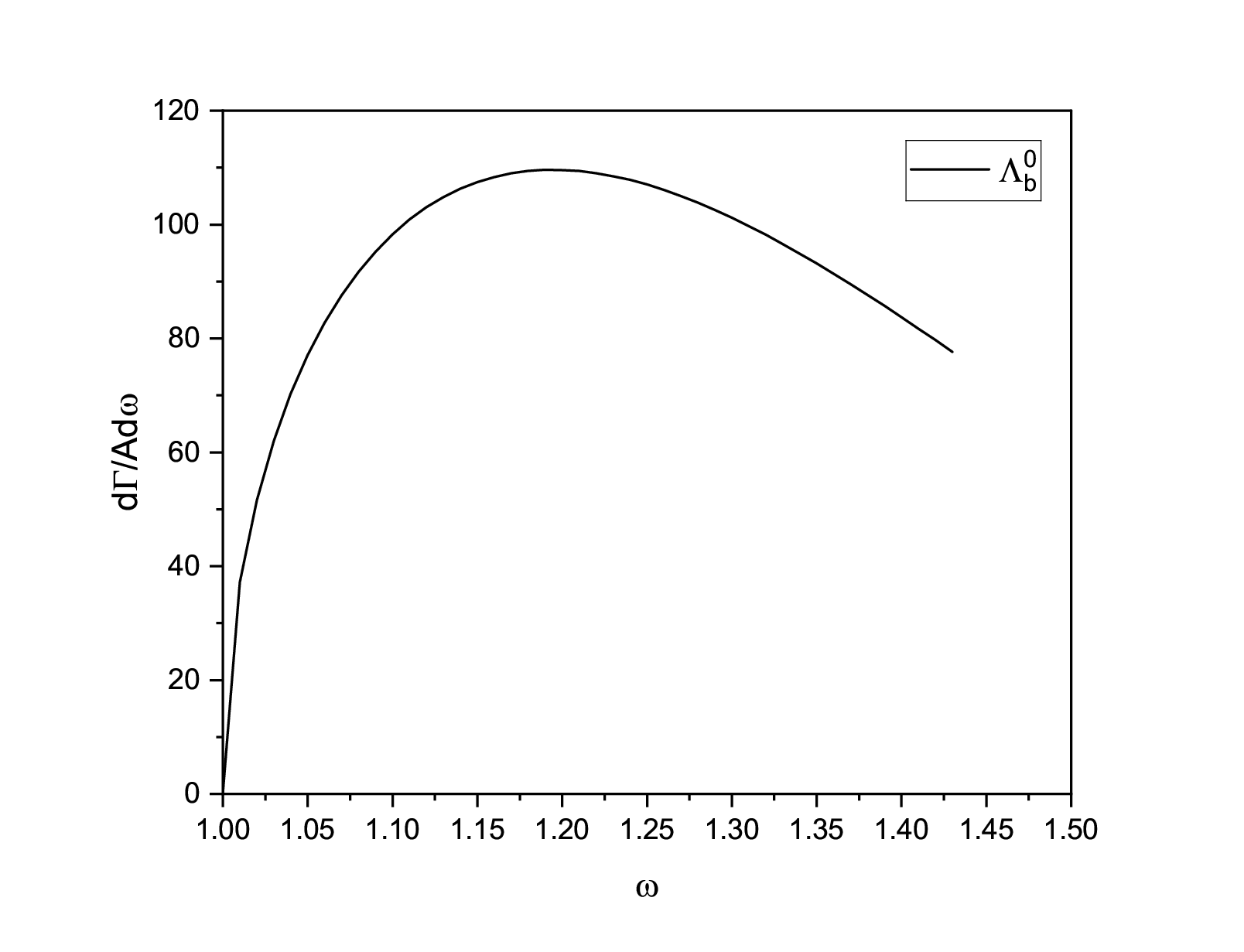}
\caption{\label{fig:7}Differential decay rates for $\Lambda_{b}^0 \rightarrow \Lambda_{c}^+ \ell\bar{\nu}$ transition}
\end{figure}
\begin{figure}
\centering
\includegraphics[scale=0.33]{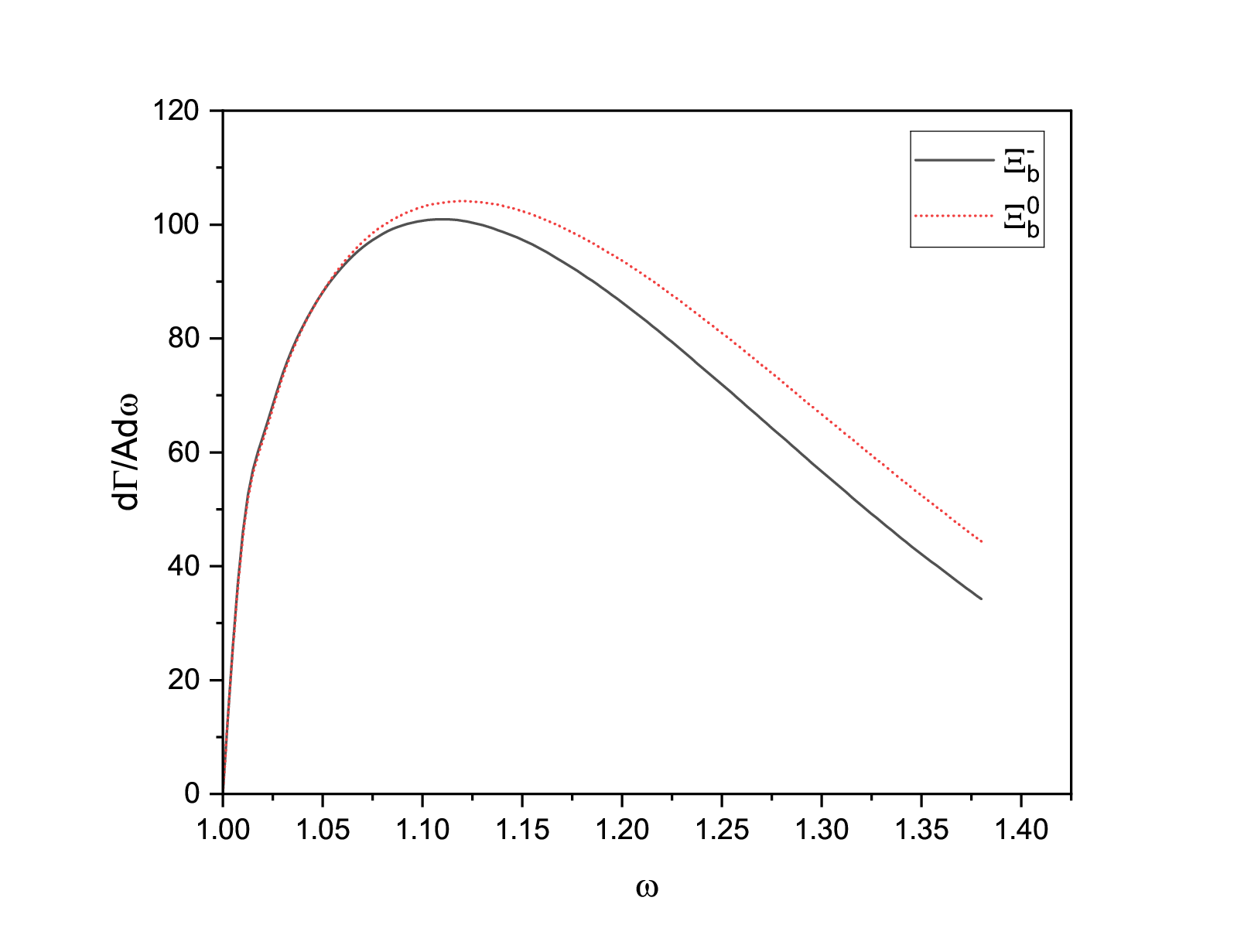}
\caption{\label{fig:8}Differential decay rates for $\Xi_{b}^{0(-)} \rightarrow \Xi_{c}^{+(0)} \ell\bar{\nu}$ transition}
\end{figure}

\begin{figure}
\centering
\includegraphics[scale=0.33]{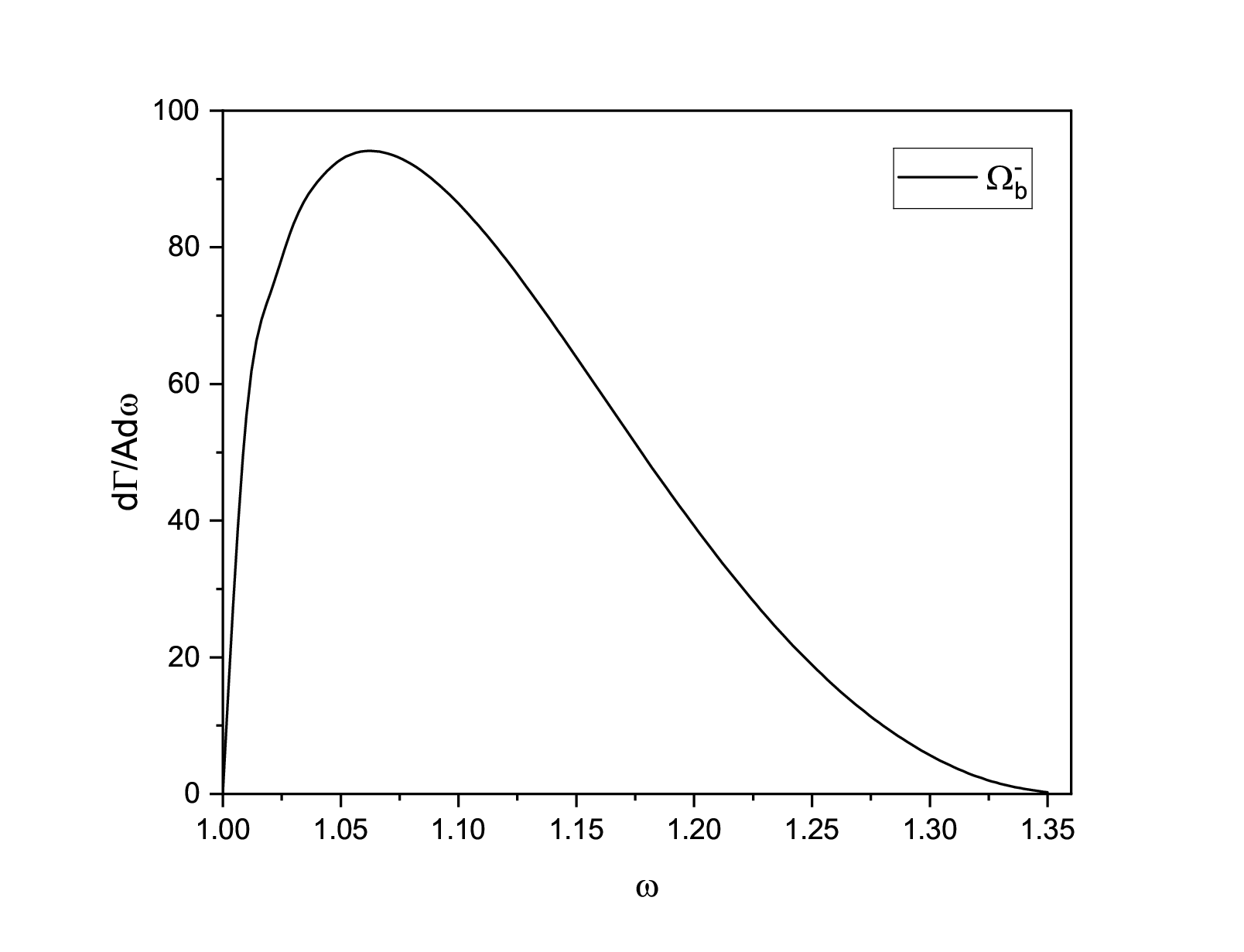}
\caption{\label{fig:9}Differential decay rates for $\Omega_{b}^- \rightarrow \Omega_{c}^0 \ell\bar{\nu}$ transition}
\end{figure}

\begin{table}[h]
    \caption{{\label{tab:table11} The slope and convexity parameter of singly bottom Baryons }}
    \begin{tabular}{ccccc}
    \hline
    Baryon	&	Slope ($\rho^2$)	&	Ref.	&	convexity parameter ($c$)	 &	 Ref.	\\
    \hline													
    $\Sigma_b^+$	&	1.02	&	2.09\cite{Ke2012}	&	0.33	&	 1.84\cite{Ke2012}		\\
    $\Sigma_b^0$	&	1.11	&	-	&	0.37	&	-		\\
    $\Sigma_b^-$	&	1.22	&	-	&	0.42	&	-\\
    $\Lambda_b^0$	&	1.12	&	$1.1\pm1.0$ \cite{Bowler1998}, 1.28 \cite{Ghalenovi2022}, 1.51 \cite{Ebert2006}	&	0.37	&	2.03 \cite{Ebert2006}		 \\
    $\Xi_b^0$	&	1.82	&	$1.4\pm0.8$ \cite{Bowler1998}, 2.27 \cite{Ebert2006}	 &	0.75	&	3.87 \cite{Ebert2006}	\\
    $\Xi_b^-$	&	1.99	&	-	&	0.86	&	-			\\
    $\Omega_b^-$	&	3.43	&	-	&	1.87	&	-		\\
    \hline
    \end{tabular}
\end{table}

\begin{table*}[h]
    \caption{\label{tab:table12} Semileptonic decay width of singly bottom Baryons (in $10^{10} s^{-1}$)}
    \tiny
    \begin{tabular}{cccccccccccccc}
    \hline
    Transition	&	Our	&	\cite{Hassanabadi2014}	&	\cite{Ivanov1997}	&	 \cite{Singleton1991}	&	\cite{Farhadi2023}	&	\cite{Korner1994}	&	 \cite{Efimov1992}	 &	 \cite{Ebert2006}	&	\cite{Ivanov1999}	&	 \cite{Rusetsky1997}	&	\cite{Ke2019}	&	\cite{Hassanabadi2015}	&	 Other Ref.	\\
    \hline
    $\Sigma_b^+\rightarrow \Sigma_c^{++}\ell\bar{\nu}$	&	6.60	&	-	&	2.23	 &	 4.3	&	$4.10\pm0.19$	&	-	&	5.4	&	1.44	&	1.9	&	2.47	&	 1.56	&	 3.04	 &	1.60\cite{Ke2012}, 1.56 \cite{Sheng2020}	\\
    $\Sigma_b^0\rightarrow\Sigma_c^+\ell\bar{\nu}$	&	6.34	&	-	&	-	&	-	 &	 -	&	-	&	-	&	-	&	-	&	-	&	-	&	-	&	-	\\
    $\Sigma_b^-\rightarrow\Sigma_c^0\ell\bar{\nu}$	&	5.99	&	-	&	-	&	-	 &	 -	&	-	&	-	&	-	&	-	&	-	&	-	&	-	&	-	\\
    $\Lambda_b^0\rightarrow\Lambda_c^+\ell\bar{\nu}$	&	5.07	&	4.92	&	 5.39	&	 5.9	&	$5.41\pm0.27$	&	5.14	&	10.4	&	5.64	&	 6.52	&	 6.02	 &	4.22	 &	-	&	5.1 \cite{Cheng1996}, $5.08\pm1.3$ \cite{Cardarelli1999}	\\
    $\Xi_b^0\rightarrow\Xi_c^+\ell\bar{\nu}$	&	4.29	&	8.22	&	5.27	&	 7.2	&	 $7.46\pm0.38$	&	5.21	&	-	&	5.29	&	6.83	&	6.4	&	-	 &	 5.22	 &	3.91 \cite{Faustov2018}	\\
    $\Xi_b^-\rightarrow\Xi_c^0\ell\bar{\nu}$	&	3.95	&	-	&	-	&	-	&	 -	&	 -	&	-	&	-	&	-	&	-	&		&	-	&	-	\\
    $\Omega_b^-\rightarrow\Omega_c^0\ell\bar{\nu}$	&	2.31	&	1.55	&	1.87	 &	 5.4	&	$1.82\pm0.12$	&	1.52	&	-	&	1.29	&	2.05	&	 2.62	 &	-	 &	 -	 &	$1.10^{+0.66}_{-0.52}$ \cite{Neishabouri2024}, 1.295 \cite{Sheng2020} \\
    \hline
    \end{tabular}
\end{table*}

\begin{table*}[h]
    \caption{{\label{tab:table13} Branching ratio calculated from Table \ref{tab:table12} using experimental mean lifetimes (in \%) }}
    \footnotesize
    \begin{tabular}{cccccccccccccc}
    \hline
    Baryon	&	Our	&	\cite{Hassanabadi2014}	&	\cite{Ivanov1997}	&	 \cite{Singleton1991}	&	\cite{Farhadi2023}	&	\cite{Korner1994}	&	 \cite{Efimov1992}	&	 \cite{Ebert2006}	&	\cite{Ivanov1999}	&	 \cite{Rusetsky1997}	&	\cite{Ke2019}	&	\cite{Hassanabadi2015}	\\
    \hline																									
    $\Lambda_b^0\rightarrow\Lambda_c^+$	&	8.25	&	7.2324	&	7.9233	&	 8.673	 &	7.9527	&	7.5558	&	15.288	&	1.47	&	9.5844	&	 8.8494	&	6.2034	 &	 -	 \\
    $\Xi_b^0\rightarrow\Xi_c^+$	&	6.35	&	12.1656	&	7.7996	&	10.656	 &	 11.0408	&	7.7108	&	-	&	7.8292	&	10.1084	&	9.472	&	-	 &	7.7256	 \\
    $\Xi_b^-\rightarrow\Xi_c^0$	&	6.21	&	-	&	-	&	-	&	-	&	 -	&	 -	&	-	&	-	&	-	&	-	&	-	\\
    $\Omega_b^-\rightarrow\Omega_c^0$	&	3.79	&	2.542	&	3.0668	&	 8.856	 &	2.9848	&	2.4928	&	-	&	2.1156	&	3.362	&	4.2968	 &	-	&	-	 \\
    \hline
    \end{tabular}
\end{table*}

\section{Results and Discussions}\label{sec:4}
The masses for $J^P=\frac{1}{2}^+$ and $J^P=\frac{3}{2}^+$ SHBs are computed within the Hypercentral Constituent Quark Model (hCQM). We adjust the model parameters $A$ and $V_0$ to get the ground state masses. Tables \ref{tab:table4} and \ref{tab:table5} summarise the results, compared with the experimental data \cite{pdg2024} and predictions from other theoretical approaches such as Lattice QCD \cite{Brown2014} and Quark diquark Model \cite{Farhadi2023}. The model quark masses and other parameters used for the calculations are listed in Table \ref{tab:table2}. The calculated masses agree with the experimental data Ref. \cite{pdg2024} and other theoretical predictions. For $\Omega_b^-$, $\Sigma_c^{*}$s, $\Lambda_c^{+*}$ and $\Omega_c^{0*}$ baryons, our calculated masses are slightly lower than the experimental values. \\
The magnetic moments for all the singly heavy baryons are computed using the spin-flavour wave functions and effective quark masses of the corresponding baryon. The calculated magnetic moments are compared with other theoretical prediction (see Tables \ref{tab:table6}, \ref{tab:table7}, and \ref{tab:table8}). We have seen good agreement for calculated magnetic moments while comparing with other theoretical approaches, particularly with Ref. \cite{Barik1983}. The predicted magnetic moment from the Bag model in Ref. \cite{Bernotas13} are comparatively lower than all other predictions. The computed magnetic moments for $J^P = \frac{3}{2}^+$ baryons (see Table \ref{tab:table8}) are in agreement with other approaches. The choice of wave functions and state mixing effects may be the main causes of the minor discrepancies among the various predictions by the different theoretical models for the magnetic moments of SHBs. The expressions of the transition magnetic moments, derived from the spin-flavour wave functions for the SHBs are listed in Table \ref{tab:table9}. Our results for the transition magnetic moments are consistent with existing theoretical models. Over all, our predicted values for the transition magnetic moments are in agreement with other theoretical predictions.

The radiative $M1$ decay widths depend upon the magnitude of the transition magnetic moment and the photon momentum $k$, the theoretical predictions follow a similar trend as the computed transition magnetic moments. The radiative $M1$ decay widths of SHBs are listed in Table \ref{tab:table10}. Currently, the experimental data for radiative decay widths of the SHBs are not available. However, there exist several theoretical estimates for the radiative decays. While comparing with other models, we find that different approaches lead to different results. For the radiative $M1$ transitions of $\Sigma_b^{0*}$, $\Sigma_c^{+*}$ and $\Omega_b^{-*}$ baryons, almost all the models predict near-zero values. For the $\Xi_c^{+*} \rightarrow \Xi_c^+\gamma$ transition, our prediction is in agreement with the EMS prediction \cite{Hazra2021} and Bag model \cite{Simonis2018}, while the pion mean-field approach \cite{Yang2020} and HBChPT \cite{Wang2019} report lower values. The reason for the variations in the predicted values of radiative decay widths by different theoretical approaches may be due to the differences in how models incorporate quark binding effects, spin-dependent interactions and relativistic corrections. Future experimental efforts on the SHBs can resolve these discrepancies among different model predictions of radiative decay widths.

We determine the IWF at zero recoil point ($\omega=1$) using Eqn. (\ref{eq:23}), where $\rho^2$ and $c$ are the slope and convexity parameter of IWF. The calculated slope and convexity parameters are listed in Table \ref{tab:table11}. Our predicted values of slope for $\Lambda_b^0$ and $\Xi_b^0$ baryons are in agreement with Lattice QCD \cite{Bowler1998}. The present computed value of slope for $\Lambda_b^0$ is 1.12 while the experimental value reported by LHCb collaboration \cite{Aaij2017} which is $1.63\pm0.07\pm0.08$. The convexity parameters for $\Sigma_b^+$, $\Lambda_b^0$ and $\Xi_b^0$ baryons, reported by others are comparatively higher than our prediction. The behaviour of the variation of IWF with respect to $\omega$ for bottom baryons are shown in Fig. \ref{fig:2} to Fig. \ref{fig:5}. At zero recoil ($\omega = 1$), we get the largest semileptonic decay rate because the initial and final baryons have the same velocities, which maximises the overlap between their wave functions. The slope is steeper for the $\Omega_b^-$ baryon than other baryons. The IWF decreases as $\omega$ increases and it is consistent for all the SHBs.

The exclusive semileptonic decay width for singly bottom baryons is listed and compared in Table \ref{tab:table12}. We have also provided the behaviour of differential decay width $\frac{d\Gamma}{Ad\omega}$ versus $\omega$. The plots for differential decay widths are shown in Fig. \ref{fig:6} to Fig. \ref{fig:9}. A good agreement of semileptonic decay width for all the baryons is observed with other theoretical predictions.

The branching ratio $Br$ is calculated using $Br = \Gamma\times\tau$, where $\tau$ is the mean lifetime of the corresponding baryon. We have calculated the branching ratio (listed in Table \ref{tab:table13}) using the semileptonic decay widths from Table \ref{tab:table12} and the experimental mean lifetime for the corresponding baryons. We haven't calculated the branching ratio for $\Sigma_b^{+,0,-}$ baryons due to the absence of experimental value of mean lifetime. In Ref. \cite{Hassanabadi2015} the branching ratio for $\Sigma_b^{+}$ baryon is calculated using the lifetime $\tau$ = $1.39 \times 10^{-12} s$. Considering $\tau_{\Sigma_b^+}$ = $1.39 \times 10^{-12} s$, our calculated branching ratio of $\Sigma_b^{+}$ baryon is 9.17\%, which is higher than the branching ratio  predicted by Ref. \cite{Hassanabadi2015} 4.22\% and \cite{Singleton1991} 5.97\%. For other baryons, we have used $\tau_{\Lambda_b}$ = $1.47 \times 10^{-12} s$, $\tau_{\Xi_b^0}$ = $1.48 \times 10^{-12} s$, $\tau_{\Xi_b^-}$ = $1.57 \times 10^{-12} s$, $\tau_{\Omega_b}$ = $1.64 \times 10^{-12} s$ \cite{pdg2024}. In Ref. \cite{Neishabouri2024} and \cite{Sheng2020} the branching ratios for $\Omega_b^-$ baryon are $1.80^{+0.99}_{-0.85}\%$ and 2.1238\%, respectively. The experimental value for $\Lambda_b^0$ baryon is $6.2^{+1.4}_{-1.3}\%$ \cite{Tanabashi2018}.
In our previous work \cite{Thakkar2020}, the value of branching ratio for $\Lambda_b^0$ baryon is 6.04\%, while in present work the value we report for $\Lambda_b^0$ baryon is 8.25\%. This variation is due to the model parameters that we have fixed. In this work, we have fixed the same model parameters for all the SHBs instead of fixing them for each baryon, while in our previous work, our focus was only on the $\Lambda_b^0$ baryon.

\section{Conclusion}\label{sec:5}
The radiative and semileptonic decays of baryons containing one heavy quark are studied within the framework of Hypercentral constituent quark model (hCQM). The ground state masses of singly bottom and singly charmed baryons are calculated by solving the six-dimensional Schr\"{o}dinger equation. The transition magnetic moments and radiative $M1$ decay widths are computed using the effective quark masses and spin-flavour wave functions. The slope and convexity parameter of the IWF are calculated at the zero recoil point. The branching ratios for singly bottom baryons are also computed and compared with other predictions. Good agreement with experimental data can be seen for masses of SHBs, confirming the reliability of the Hypercentral Constituent Quark Model for describing heavy baryon interactions.\\
\\
Data Availability Statement: No Data associated in the manuscript.

\end{document}